\documentclass[twocolumn,showpacs,showkeys,preprintnumbers,amsmath,amssymb,superscriptaddress]{revtex4}

\usepackage{amsmath}
\usepackage{upgreek}
\usepackage{latexsym}
\usepackage{bm}
\usepackage{dcolumn}
\usepackage{graphicx}
\usepackage{epsfig}

\begin{document}

\title{Bottlenecks in granular flow: \\ When does an obstacle increase the flowrate in an hourglass?} 

\author{F. Alonso-Marroquin}
\email{fernando.alonso@sydney.edu.au}
\affiliation{School of Civil Engineering, The University of Sydney, Sydney NSW 2006, Australia}

\author{S. I. Azeezullah}
\affiliation{School of Mathematics and Physics, The University of Queensland, St Lucia QLD 4072, Australia}
\affiliation{School of Civil Engineering, The University of Sydney, Sydney NSW 2006, Australia}

\author{S. A. Galindo-Torres}
\affiliation{School of Civil Engineering, The University of Queensland, St Lucia QLD 4072, Australia}

\author{L. M. Olsen-Kettle}
\affiliation{School of Earth Sciences, The University of Queensland, St Lucia QLD 4072, Australia}

\date{\today}%

\begin{abstract}
Bottlenecks occur in a wide range of applications from pedestrian and traffic flow to mineral and food processing.  We examine granular flow across a bottleneck using particle-based simulations. Contrary to expectations  we find that the flowrate across a bottleneck actually increases if an optimized obstacle is placed before it. The dependency of flowrate on obstacle diameter is derived using a phenomenological velocity-density relationship that peaks at a critical density.  This relationship is in stark contrast to models of traffic flow, as the mean velocity does not depend only on density but attains hysteresis due to interaction of particles with the obstacle.
\end{abstract}

\maketitle

The improvement in flowrate of particles passing through a bottleneck has applications ranging from industrial granular flow \cite{johanson1,johanson2} and traffic flow \cite{helbing97aac} to escape dynamics under panic \cite{helbing2000sdf}.  Optimal plant design for the conveyance and storage of powders and bulk solids is a challenge faced by nearly all industries, from powder coating to food, from nano-scale powders and pharmaceuticals to cement, coal, and ore \cite{schulze}.  Since the sixties, empirical placement of inserts (obstacles) before outlet openings has been used in silo design for a number of reasons, including: transformation of the flow profile from funnel flow to mass flow, enlargement of the mass flowrate, reduction of the stresses in the silo, and mixing and homogenizing of bulk solids \cite{ johanson1,johanson2,schulze}. Whereas the flow problems frequently found in silo flow are relatively well-known, improvements to design of optimized escape routes for panicking crowds are at their inception.  

Studies on escape dynamics under panic have shown that obstacles  placed before the outlet can lead to big changes in flow patterns. Due to the difficulties in performing real experiments with humans, simulations of self-driven particles have been proposed.  These simulations have already shown, counterintuitively,
the benefit of placing an obstacle before an exit to prevent  or reduce injuries under conditions of panicked escape \cite{helbing2000sdf,escobar2003ads}.   Placing a column in front of an exit substantially reduced evacuation time for ants squirted with citronella~\cite{shiwakoti09,burd10}.  Escobar and De la Rosa explain this behavior as the ``waiting room'' effect ~\cite{escobar2003ads} whereby particles slow down and accumulate above the obstacle, decongesting the exit and increasing the flowrate. 
We provide full modeling of the flow of particulate materials around such an obstacle where we observe a peak in flowrate for optimized conditions. We develop a consistent framework based on statistics from simulations which characterizes the relationship between the flowrate and the geometry of the bottleneck and obstacle.

We present a parametric study from statistical analysis of $6,080$ simulations of gravity-driven granular flow
through an hourglass hopper. We place a circular obstacle above the bottleneck and investigate how it affects the flowrate.   Statistical analysis of the simulations shows that for a specific range of obstacle diameters and positions the flowrate is higher than the flowrate without an obstacle. The  position and velocity of the particles are used to derive the dependency of the flowrate on the diameter of the obstacle from a multilinear relation between mean velocity, density, and obstacle diameter. We also investigate the connection between this multilinear model and velocity-density relations used in traffic flow.

\begin{figure}[t]
\includegraphics[trim = 5mm 0mm 9mm 0mm, clip, width=0.24\linewidth]{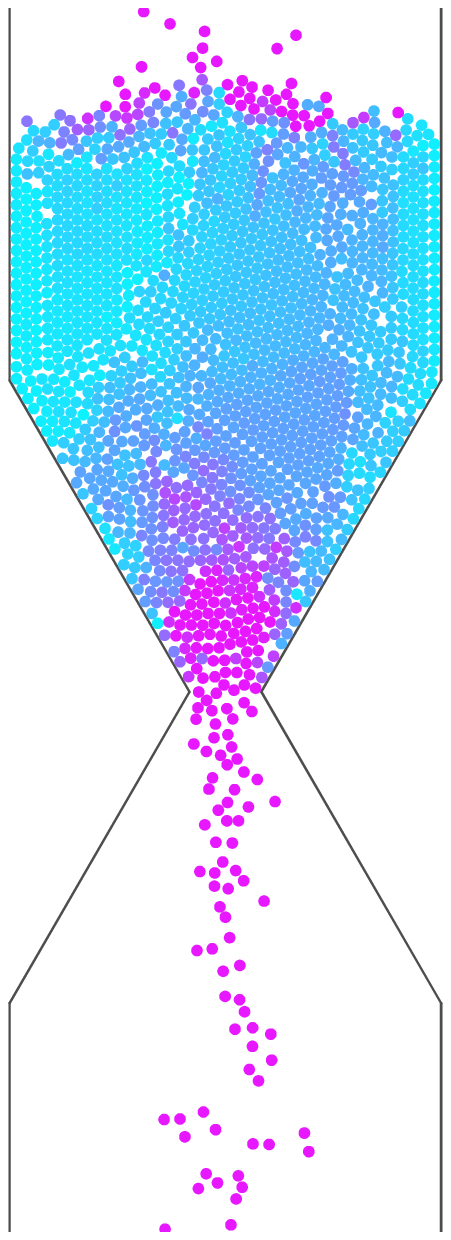}
\includegraphics[trim = 5mm 0mm 9mm 0mm, clip, width=0.24\linewidth]{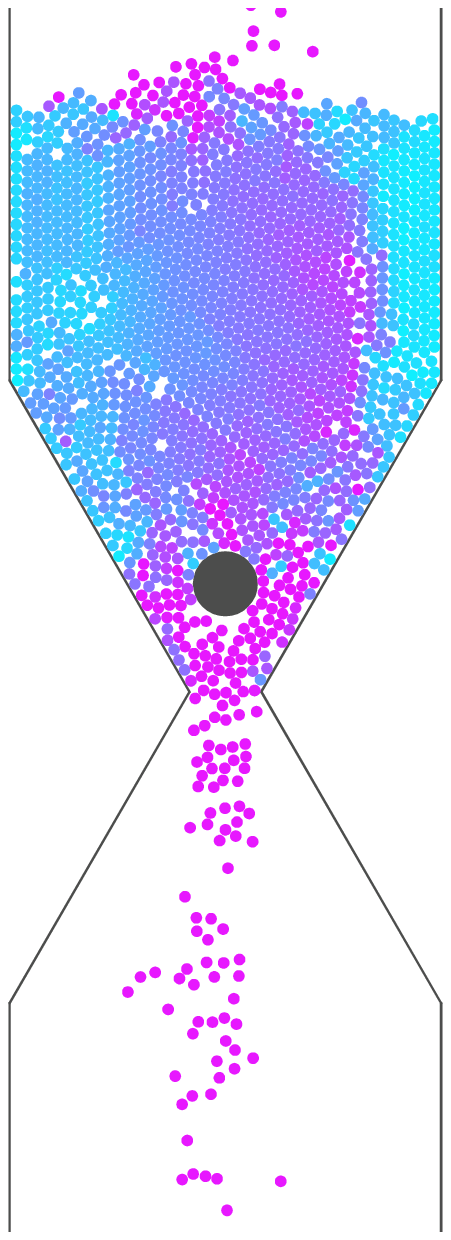}
\includegraphics[trim = 5mm 0mm 9mm 0mm, clip, width=0.24\linewidth]{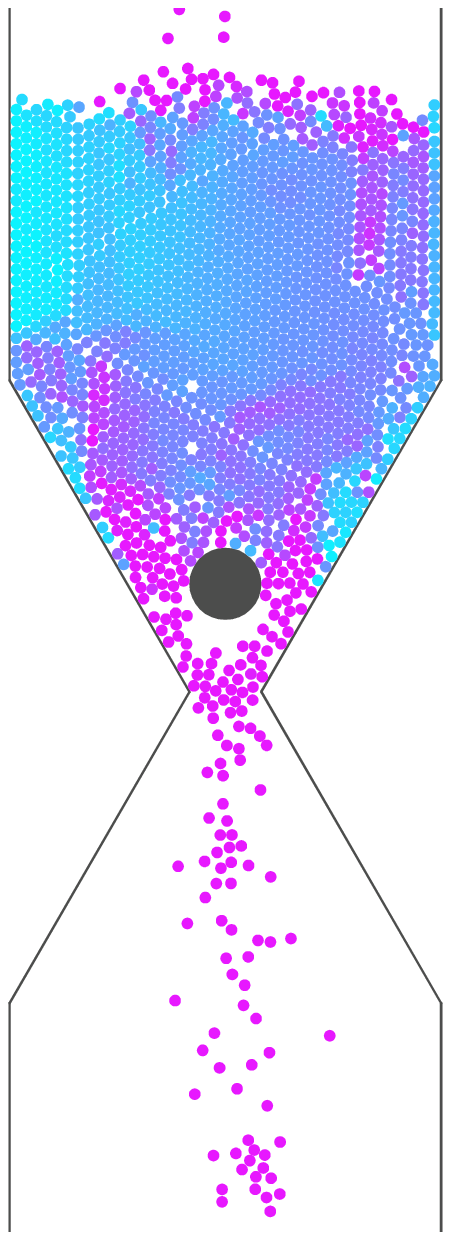}
\includegraphics[trim = 5mm 0mm 9mm 0mm, clip, width=0.24\linewidth]{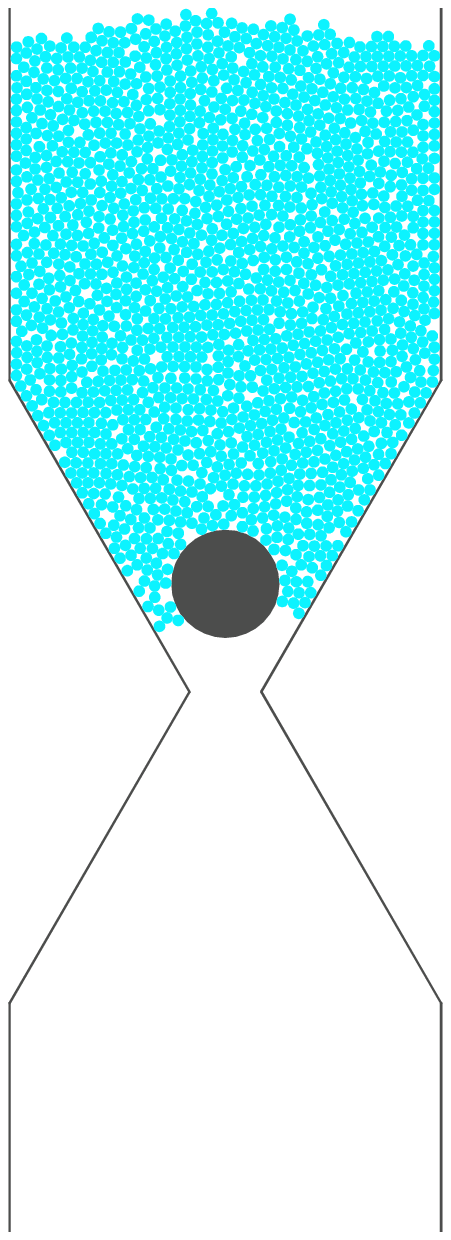}
\caption{ \label{fig:snapshots} Snapshot of the simulation of gravity-driven granular flow at a time of $20$~s with no obstacle (left), and with an
obstacle of $1.8$ m, $2$ m,  and $3$ m diameter (right). The color encodes particle speed (darker is faster). The hopper angle with respect to the vertical is $30^o$.}
\end{figure}

All simulations consider circular particles passing through a neck of width $W$ of an hourglass-shaped hopper with angle $\theta$  with respect to the vertical as shown in Fig.~\ref{fig:snapshots}.   A circular obstacle of diameter $D$ is placed centrally at a distance $H$ above the bottleneck. The particles interact with each other via elastic, viscous, and frictional forces, and are subject to gravity. Details of the particle-based model are presented in \cite{alonso08epl,alonso09}. The parameters of the model are: the normal contact stiffness $k=10^8$ N/m, the tangential contact stiffness $k_t = 0.1$ $k_n$, friction coefficient $\mu =0.2$, the normal and tangential coefficients
of viscosity  $\gamma_n=800$ s$^{-1}$  and $\gamma_t=80$ s$^{-1}$, and gravity  $g=10$ m/s$^2$. The size of the hopper is $12$~m~$\times$~$32$~m. The  particles have a diameter of $31.7$ cm and $7.89$ kg of mass.  The width of the bottleneck is fixed to $W=2$ m, which is wide enough to avoid clogging. Simulations were repeated with varying angle $\theta$, diameter $D$, height $H$ of the 
center of the obstacle above the neck. The default values are  $D= 2$ m, $H = 3$ m, and $\theta=30^o$.  Each sample runs for sixty seconds.

Initially the particles are placed at the top half of the hopper, and the obstacle is fixed above the center of the neck. 
Gravity ensures the particles flow through the bottleneck. 
We enforce periodic boundary conditions by replacing  the particles that have reached the bottom back to the top of the hopper. The average number of refilling particles per second is used to measure the flowrate.
Differences in initial conditions cause statistical fluctuations, so for each  parameter suite the simulation was repeated
with the particles in different initial positions. This yielded reliable average flowrates. The uncertainty of the measurement is estimated  from the standard error: $\epsilon = \sigma/\sqrt{s}$, where $\sigma$ is the standard deviation of the values and $s$ is the number of samples.  The default value is $s=32$ samples.

The first interesting observation from the simulations is the complex dependency of the density and velocity of the particles on the size of the obstacle.
Snapshots for different diameters at time $20$~s are shown in  Fig.~\ref{fig:snapshots}. 
For obstacle diameters less than $1.8$ m the obstacle reduces the 
velocity of the particles slightly, while it does not significantly affect the density of particles around the bottleneck. This trend changes dramatically
when the obstacle diameter lies between $1.8$ m and $2.1$ m. 
In this range the obstacle produces the waiting-room effect \cite{escobar2003ads}, with reduction of density above the bottleneck and subsequent
increase in velocity of particles, and therefore in flowrate. This waiting-room effect disappears for obstacle diameters above $2.1$ m. In this case the distance between the obstacle and the hopper walls is narrow enough to 
produce clogging, leading to a decrease in the flowrate.

\begin{figure}[t]
(a)\qquad\qquad\qquad\qquad\qquad(b)\\
\includegraphics[width=0.48\linewidth]{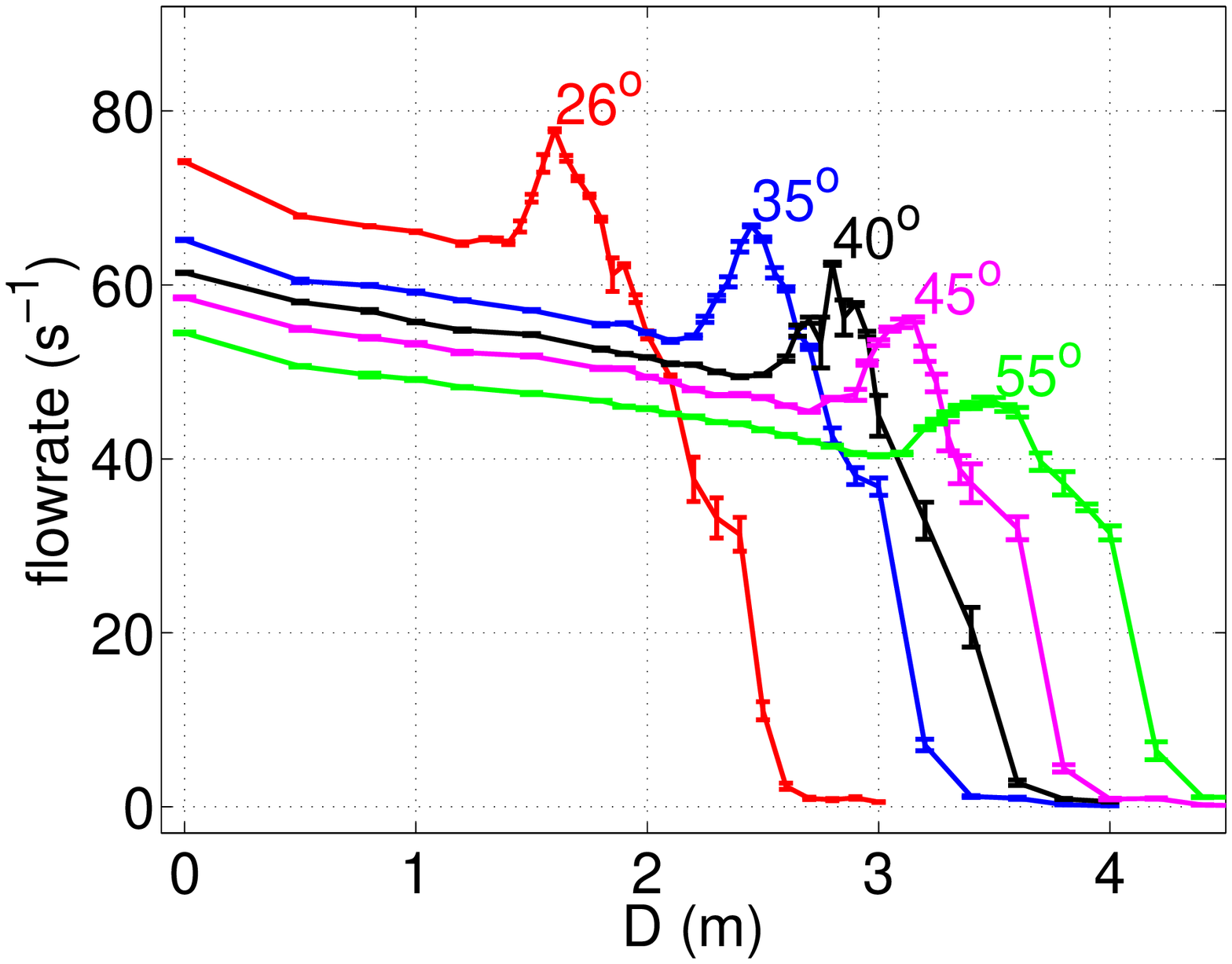}
\includegraphics[width=0.48\linewidth]{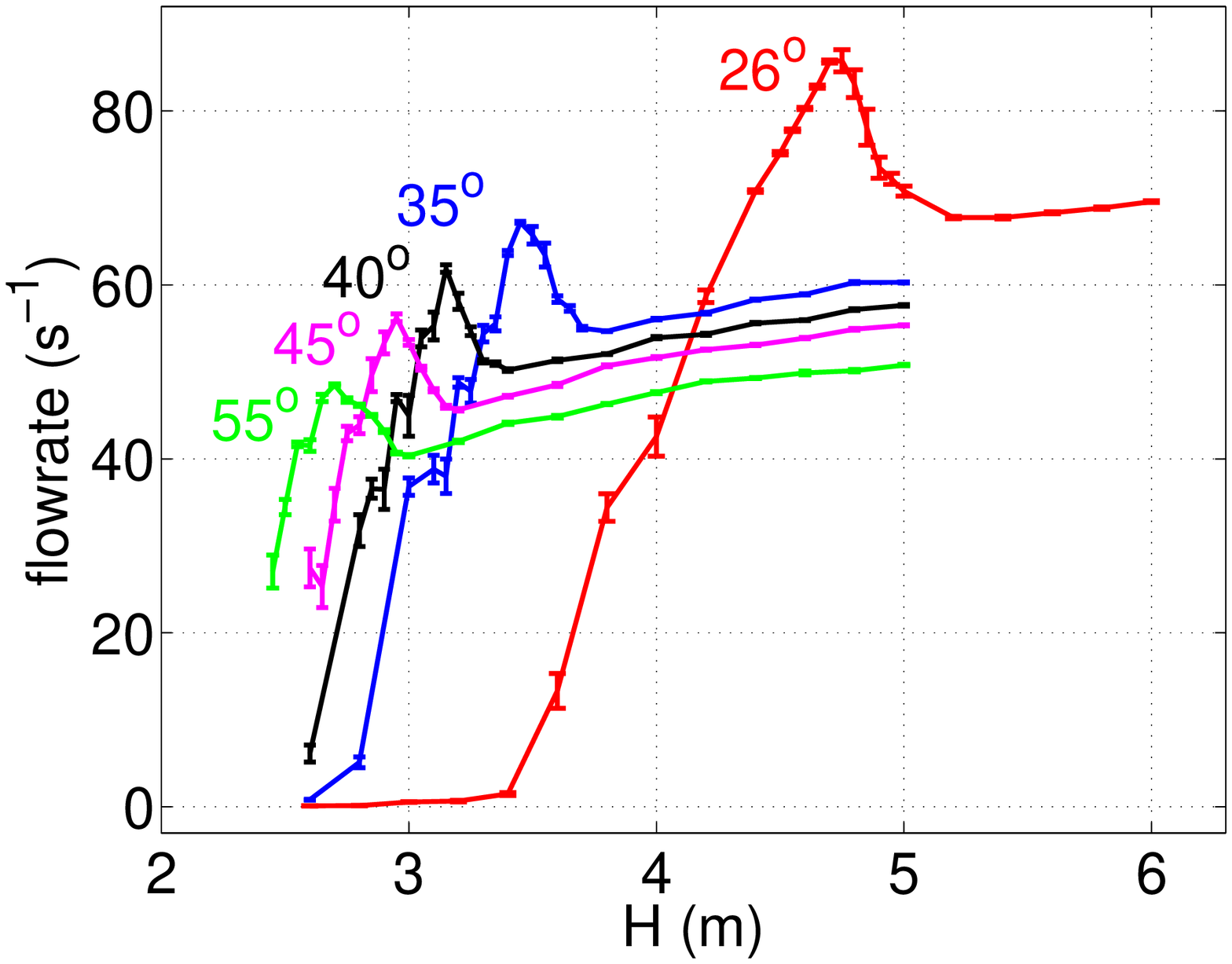}
(c)\qquad\qquad\qquad\qquad\qquad (d)\\
\includegraphics[width=0.48\linewidth]{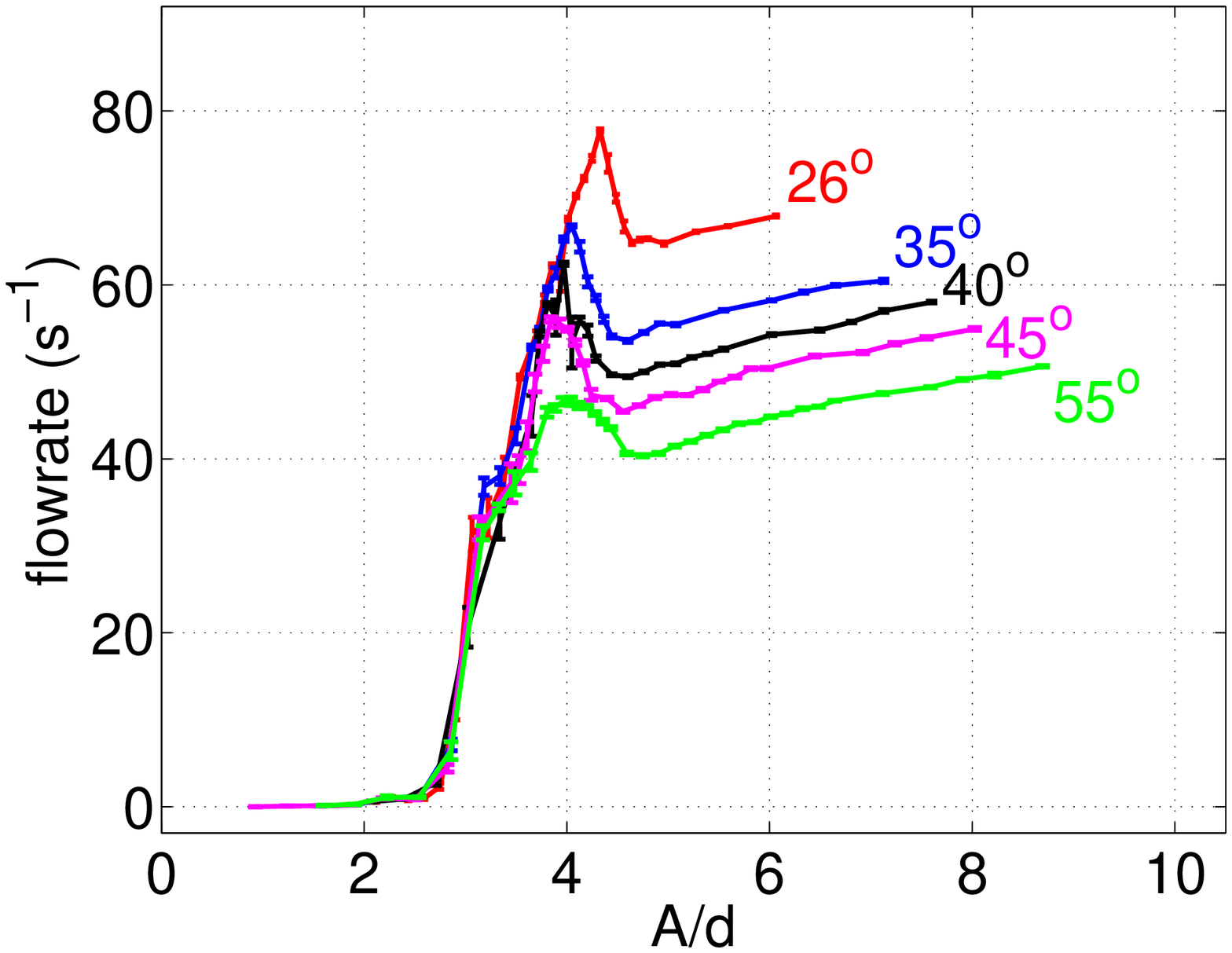}
\includegraphics[width=0.48\linewidth]{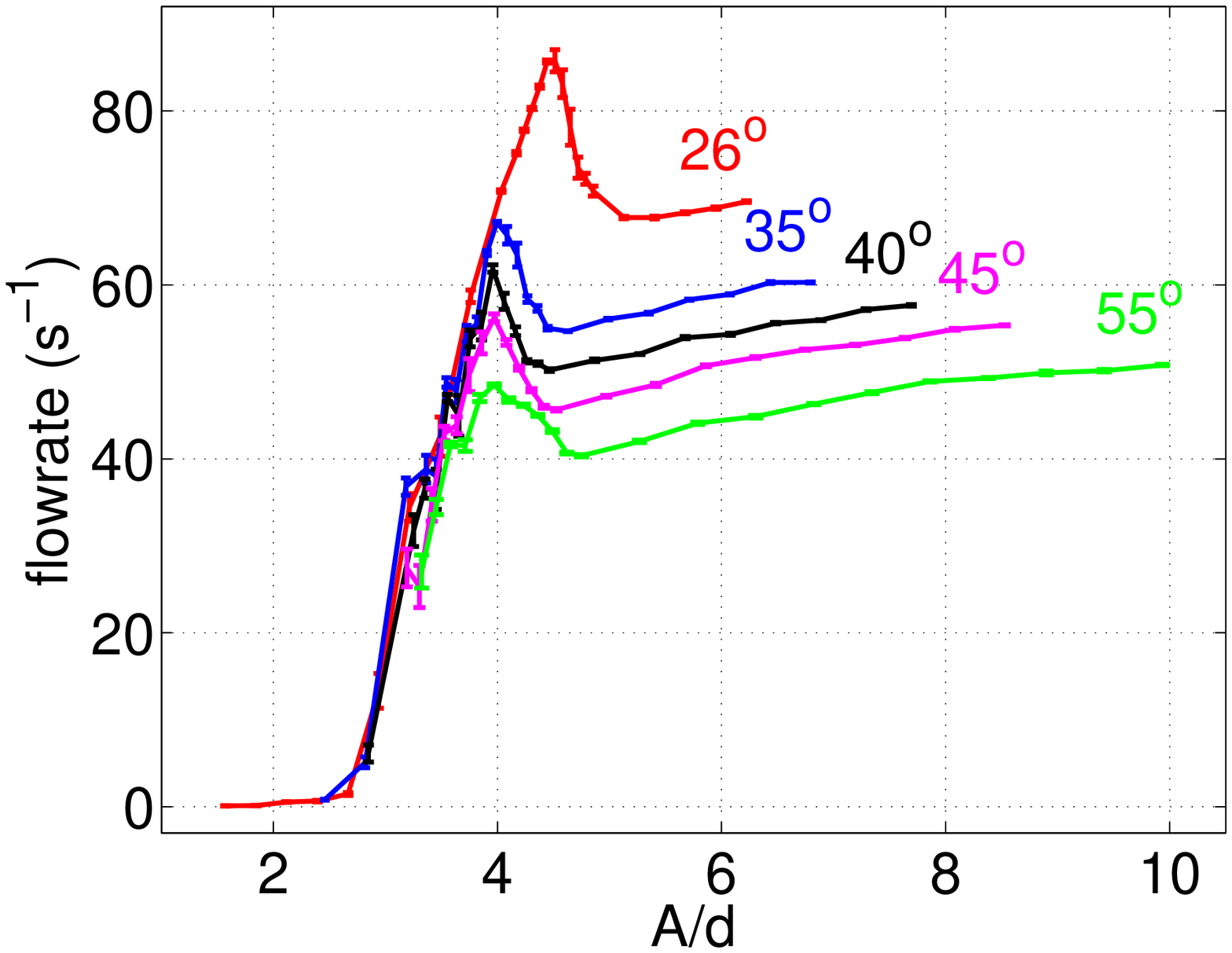}
\caption{Flowrate versus obstacle diameter in (a), and versus obstacle height in (b), for different values of hopper angle $\theta$. 
In (a) the center of the obtacle is $3$ m above the center of the neck. In (b) the obstacle diameter is $2$ m. Each simulation has $414$ particles.
(c) and (d) show the flowrate versus normalized aperture $A/d$ for (a) and (b). $A$ is the minimum distance from the obstacle to the hopper and $d$ is 
the diameter of the particles.
\label{fig:flowrate}}
\end{figure}

The first series of simulations are performed to investigate the effect of  obstacle size on the flowrate. We fixed the height of the center of the obstacle to 
$3$ m above the center of the bottleneck and the number of particles to $414$. 
Fig.~\ref{fig:flowrate} (a) shows the flowrate vs.~obstacle diameter $D$ for different hopper angles $\theta$.  
For each $\theta$ the flowrate peaks at a finite value of obstacle diameter $D_c(\theta)$ that we call the optimal diameter.  As the angle of the 
hopper increases from $26^o$ to  $55^o$ the optimal diameter increases while the peak flowrate decreases. An important control parameter
for the flowrate is the aperture, which is defined as the minimum distance between the obstacle and the hopper.  Fig.~\ref{fig:flowrate} (c) 
shows the flowrate vs.~this aperture normalized by the diameter of the particles. We observe a slight dependence of the optimal aperture to the hopper angle.
The optimal aperture is around $4.1(2)$ particle diameters for all values of $\theta$.

In the second series of simulations we investigate how the flowrate depends on the height of the obstacle. We fixed the diameter of the obstacle to $2$ m and varied the height $H$ and angle $\theta$.  The results are shown in Fig.~\ref{fig:flowrate}~(b). For each hopper angle there is an optimal height for which the flowrate is maximal. Again as the angle of the hopper increases the peak decreases and the optimal height becomes smaller.  The flowrate is plotted against the normalized aperture in Fig.~\ref{fig:flowrate}~(d), and  we find curves similar to the case (c) when the height is fixed and the diameter is varied. In both cases the optimal aperture is around $4.1(2)$ particle diameters. The main difference is that when the height of the obstacle is varied a higher flowrate is reached than when changing its diameter. For example for a hopper angle of $26^o$ the maximal flowrate for optimal diameter is $77.8\pm 0.3$ particles per second, while it reaches $85.7\pm 0.1$ particles per second when the height is optimized. In comparison these flowrates are both greater than the flowrate without obstacle, which is $74.2\pm 0.1$ $s^{-1}$. The difference between the flowrate-aperture curves in  Fig.~\ref{fig:flowrate} (c) and (d) shows that the flowrate depends not only on the aperture, but also on the area enclosed between the obstacle and the bottleneck.

\begin{figure}[t]
(a)\qquad\qquad\qquad\qquad\qquad(b)\\
\includegraphics[width=0.48\linewidth]{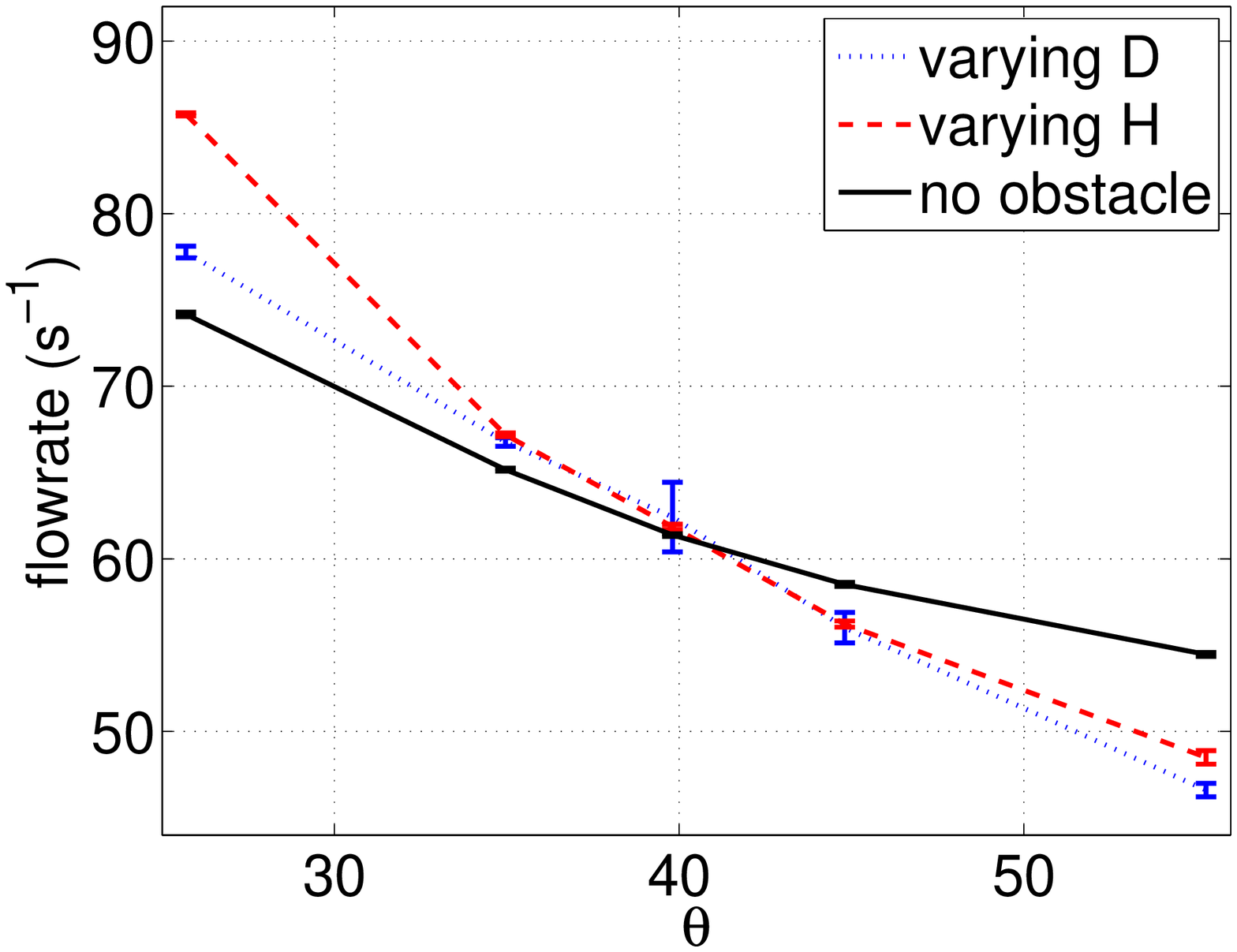}
\includegraphics[width=0.48\linewidth]{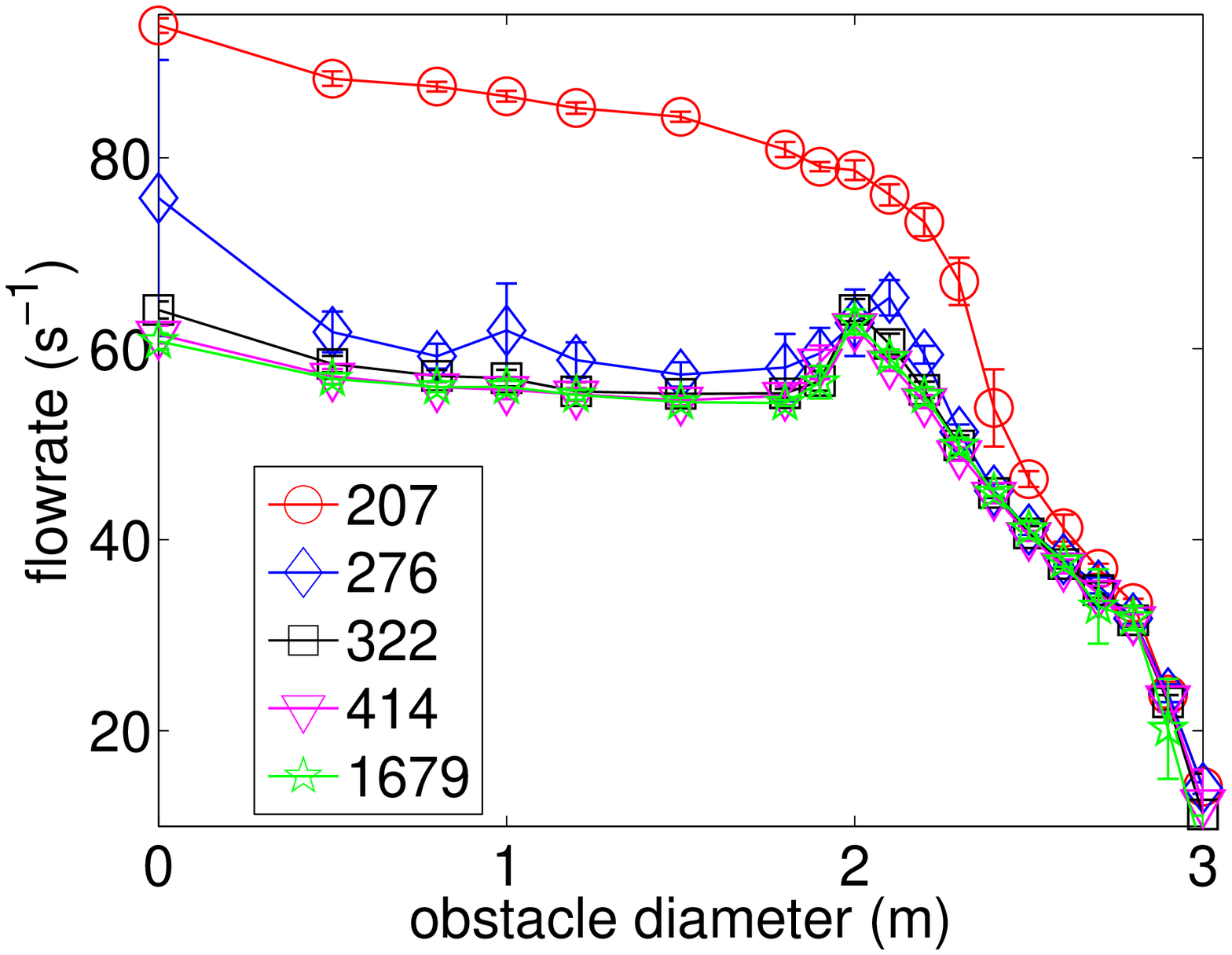}
\caption{ In (a), flowrate versus hopper angle $\theta$ is shown for 3 cases: optimized obstacle with $D$ varied and $H= 3$~m; 
optimized obstacle with $H$ varied and $D = 2$~m; and no obstacle. The flowrates are calculated by averaging over 
$32$ samples of $414$ particles each. In (b), flowrate versus obstacle diameter is shown for $5$ different numbers of particles, 
averaging over $8$ samples in each case.}
\label{fig:FvsTheta}
\end{figure}

In Fig.~\ref{fig:FvsTheta} (a) we compare the flowrate with optimized obstacle with the flowrate with no obstacle for different values of hopper angle. When we vary $H$ we employ a constant $D=2$ m, and when we vary $D$ we employ a constant $H=3$ m.
When the hopper angle is below $40^o$ the flowrate with an optimized obstacle is larger than without an obstacle. The main conclusion of these 
two series of simulations is that for hopper angles below $40^o$ there is a narrow range of apertures and obstacle positions where the flowrate is higher than the corresponding flowrate without an obstacle.  More support and evidence for the optimized flowrate due to obstacle placement before an outlet opening can be found from numerical simulations of panic-driven particles \cite{helbing2000sdf}, experiments with ants \cite{burd10}, and more practically in cattle herding where the stockman has the role of the obstacle \cite{shiwakoti09}.  The contribution of our simulations is to show that obstacle placement needs to be optimized for improvement in the flowrate. More specifically,  there is only a very narrow, specific range of parameters where the flowrate is higher than that without an obstacle.

\begin{figure}[t]
\begin{center}
(a)\qquad\qquad\qquad\qquad\qquad(b)\\
\includegraphics[width=0.48\linewidth]{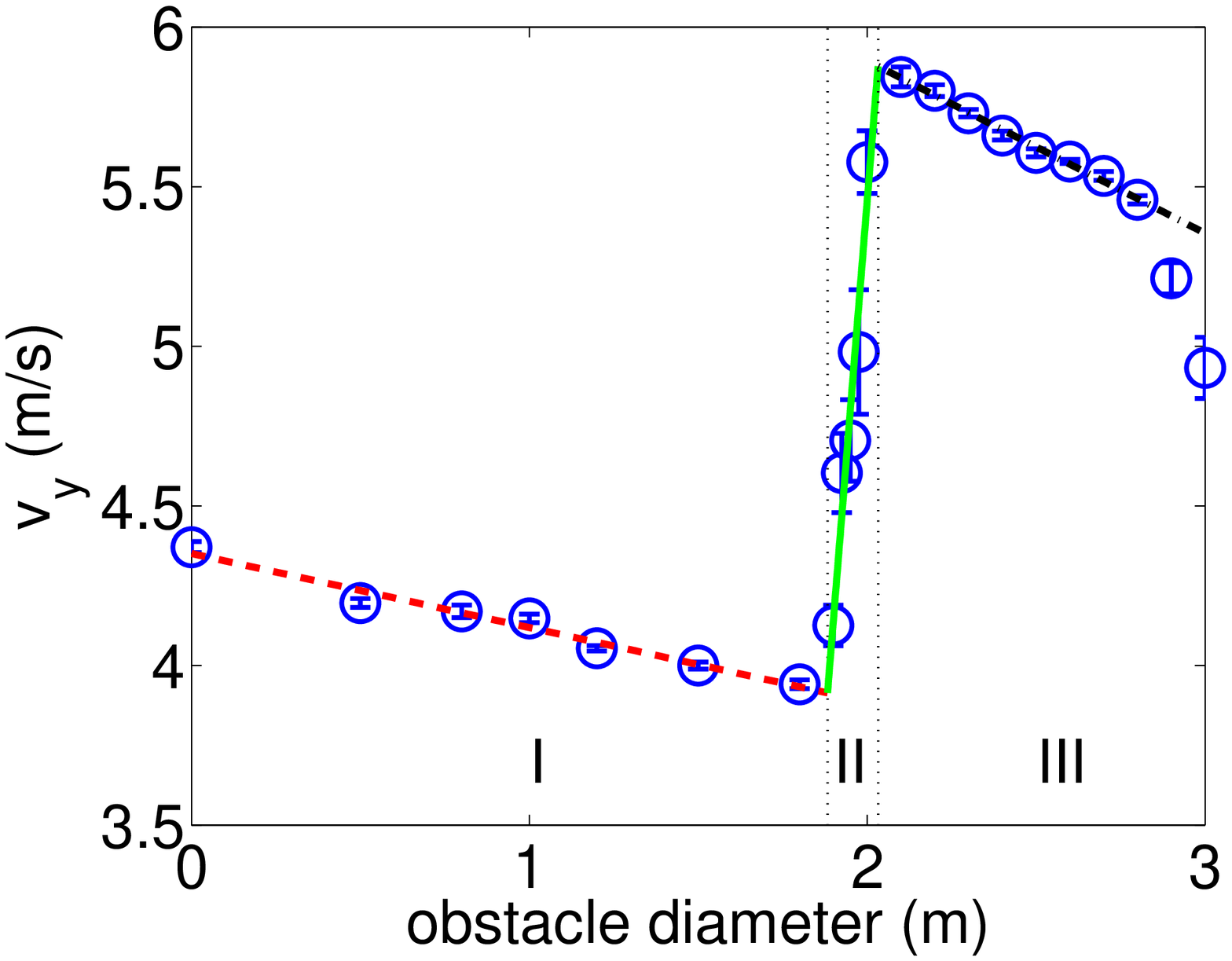}
\includegraphics[width=0.48\linewidth]{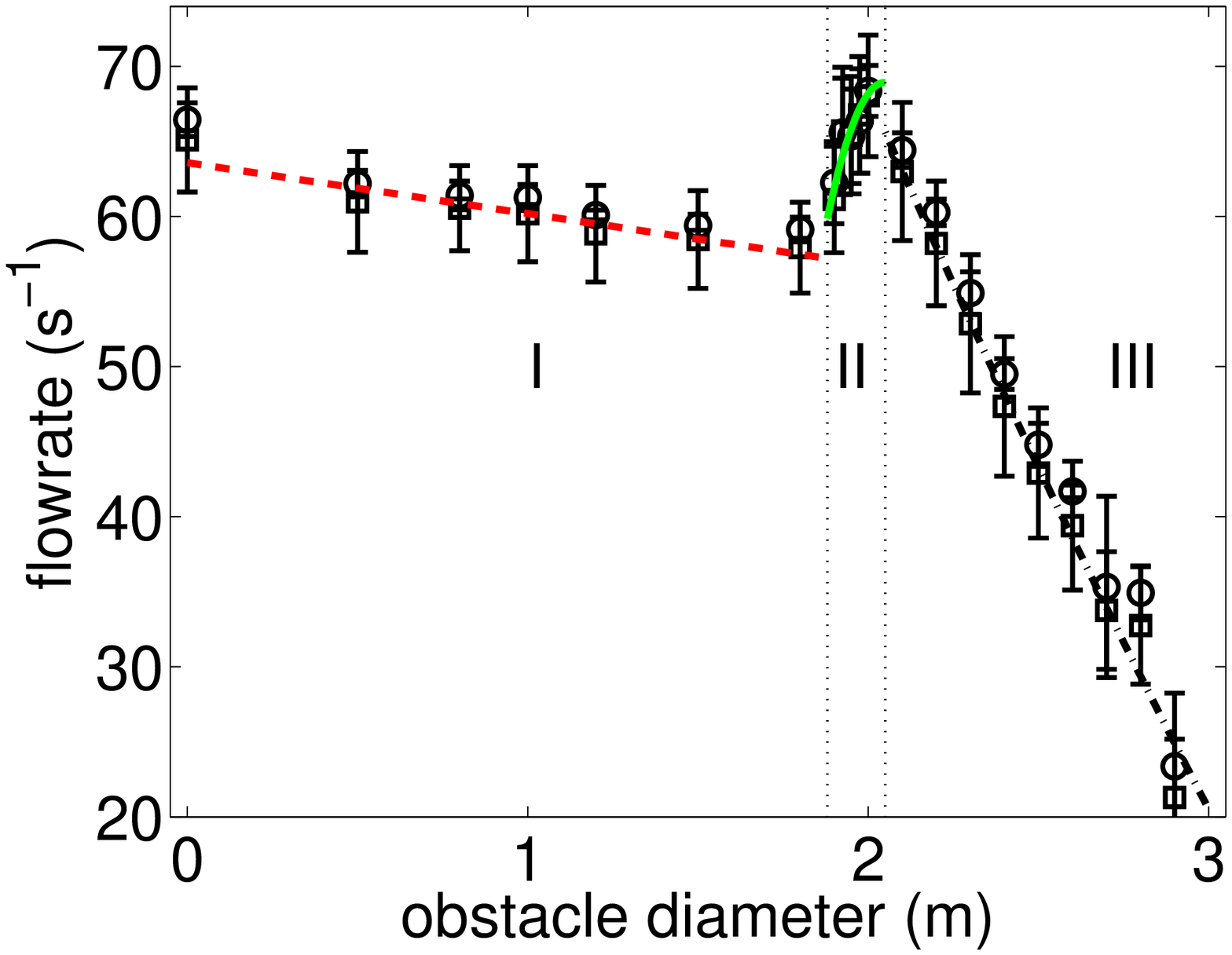}
(c)\qquad\qquad\qquad\qquad\qquad(d)\\
\includegraphics[width=0.48\linewidth]{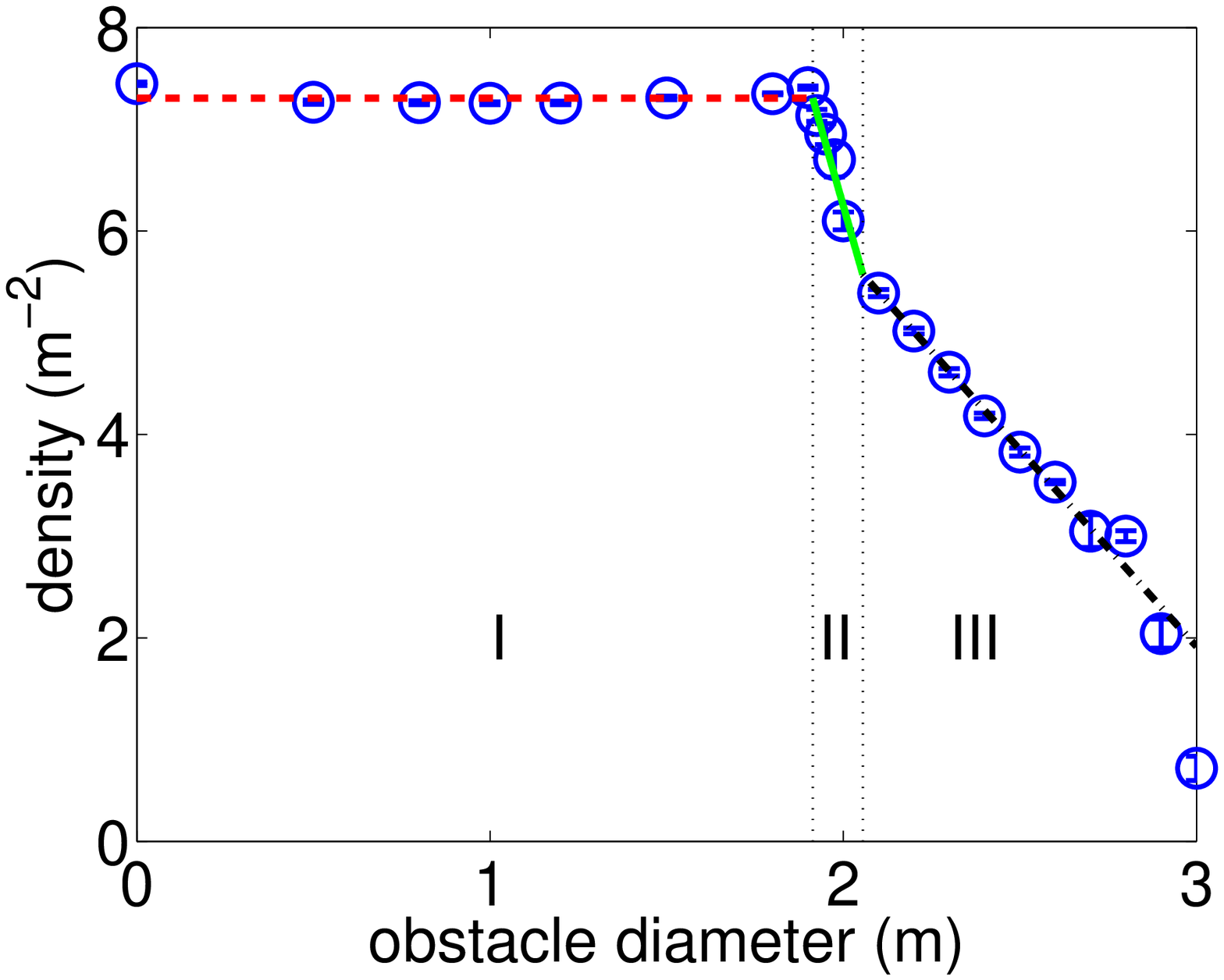}
\includegraphics[width=0.48\linewidth]{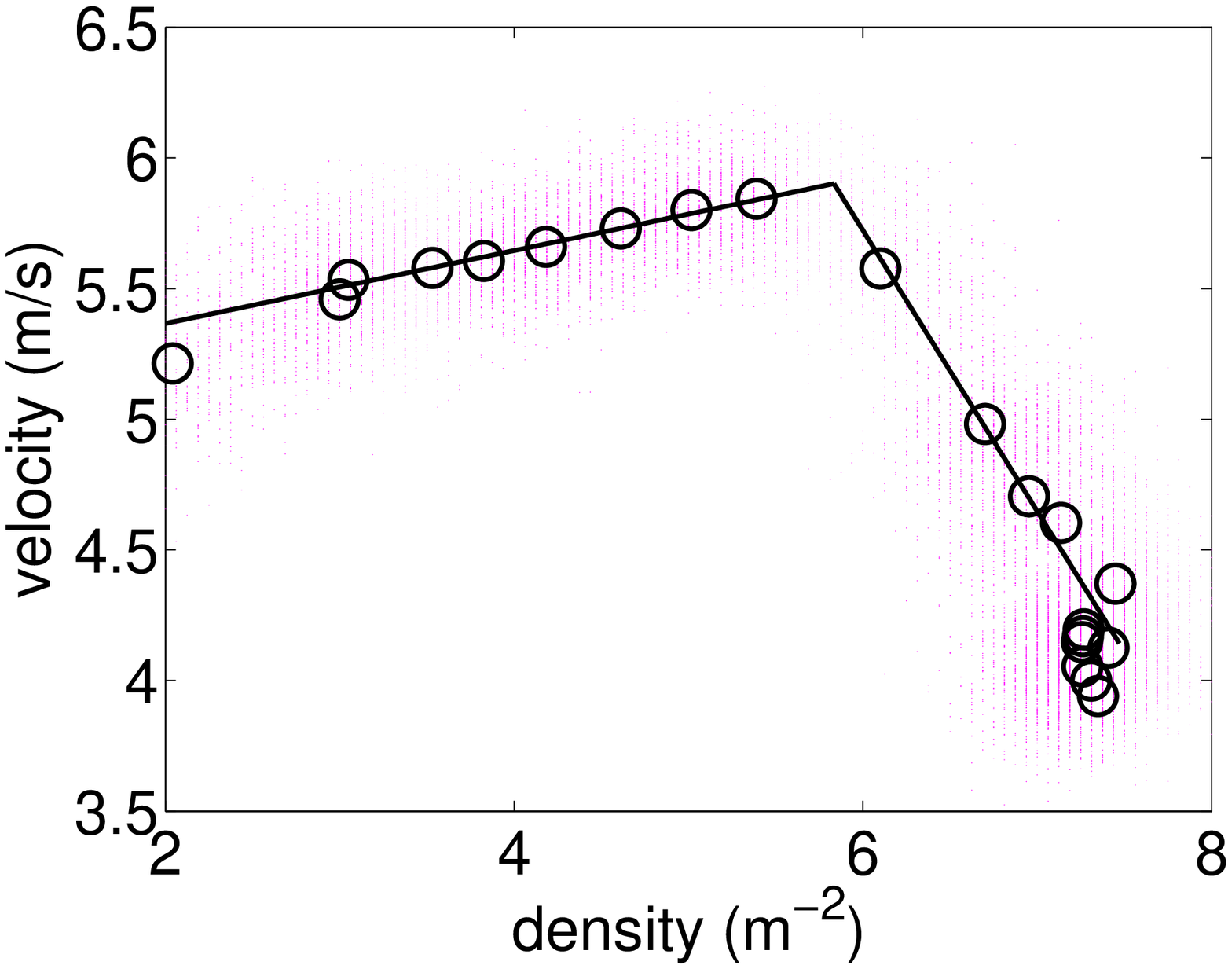}
\end{center}
\caption{\label{fig:flowrate2} 
(a) Vertical velocity at the bottleneck versus obstacle diameter. (c) Density at the bottleneck versus obstacle diameter.
The lines in (a) and (c) show the multilinear fit of the data. The squares in (b) show the flowrate versus obstacle diameter
calculated as the number of refilling particles per second. The circles in (b) show the flowrate when calculated from the
hydrodynamic relation $J=\rho VW$. The lines in (d) correspond to the model using the fitting in (a) and (b). Each dot in (d) shows the velocity versus density in a snapshot of the simulation. The circles show the time average of 
both density and velocity at different obstacle diameters. The bi-linear fitting is performed using the slopes
obtained from zones II and III in (a) and (c). 
}
\end{figure}


It appears that flowrate depends strongly on two geometric parameters: aperture and hopper angle, and only weakly on
number of particles. Fig.~\ref{fig:FvsTheta}~(b) shows that for $207$ particles the flowrate does not improve at a critical obstacle diameter since they rarely collide in the bottleneck; from $276$ to $414$ particles
the flowrate decreases slightly; and with more than $414$ particles there is no dependency on the number of particles.


We now construct a microscopic traffic-flow model to describe the phenomenology of the
flowrate. Traffic-flow modeling assumes that in the steady state the mean velocity $V$ 
of the particles (car or pedestrians) is a  function of the density $\rho$ 
\cite{helbing97aac,treiber1999dpa}. The flowrate density (also called the hydrodynamic relation) is just the product of the density and 
the mean vertical velocity, $J=\rho V$,  which implies a dependence of the flow 
solely on the density. 

To obtain the velocity-density relation for our problem, we focused on the rectangular area $5$ cm above and
below the bottleneck. The density is calculated as the number of particles in this rectangle divided by its area.
The mean velocity is obtained by averaging the vertical velocity of the particles within this rectangle. 
Time averages of these two quantities are obtained from $800$ snapshots of  the simulation.  
In this analysis we use $2185$ particles and fix the hopper angle to $30^o$. 

Fig.~\ref{fig:flowrate2} (a) and (c) show the density and mean velocity versus obstacle 
diameter averaged over $800$ snapshots and eight samples.  Depending on the obstacle diameter 
we can distinguish three flow regimes: 
{\bf Zone I} corresponds to obstacle diameters below $1.8$ m. In this regime
the averaged velocity decreases as the diameter of the obstacle increases, while the density
keeps almost constant. We call it a {\it hysteresis} zone, because there the mean velocity
does not depend on density but on the history of particles. Here  the particles crossing the bottleneck ``remember" 
whether they collided with the obstacle because if they did their velocity is lower than the one if they had not collided.
{\bf Zone II} corresponds to an obstacle diameter of $1.8-2.1$ m. This is the most fascinating regime, where
we observe a decrease in the density as well as an abrupt increase in velocity.
This narrow zone is where the waiting-room effect takes place leading to high flowrates.
We call this regime the {\it congested zone}, because it resembles the regime in vehicular flow 
where a decrease in the density of cars leads to an increase of the mean velocity of the cars.
{\bf Zone III} corresponds to obstacle diameters above $2.1$ m.  In this zone the region between 
the bottleneck and the obstacle is  loosely packed and  the particles are relatively  free and unconstrained.
The transition from zone II to zone III  is characterized  by a change in the relationship between the velocity 
and density. While the density is still  decreasing  with an increase of obstacle diameter, the velocity in this zone changes from increasing
to decreasing. In zone III the particles behave similarly to the {\it free-flow} regime in traffic flow, as the reduction in density of 
cars leads to a  reduction of flowrate (see Fig.~\ref{fig:flowrate2} (a) and (c)).  

We note a significant difference between zone III and the free-flow regime in traffic-flow models \cite{helbing97aac,treiber1999dpa}.
In our simulations the velocity decreases when density decreases, while the traffic-flow model assumes that the velocity
decreases when density increases.   The transition from zone II to zone III in Fig.~\ref{fig:flowrate2} (a)  is related to onset 
of clogging in the apertures between the obstacle and the hopper walls. This clogging makes the mean velocity at the
bottleneck turn from increasing to decreasing as the diameter is increased further. Clogging means that the particles are slowing 
down in the apertures and that fewer particles are getting through the apertures leading to a both a decrease in density and velocities 
in the bottleneck. Thus the velocity is decreasing with decreasing density in the bottleneck for zone III.

The three regimes can be fitted using a multilinear relation of the velocity and density versus 
obstacle diameter. This relation is then used to calculate the dependency of flowrate on
obstacle diameter, as shown in Fig.~\ref{fig:flowrate2} (b). We observe from this multilinear
model that the optimal diameter corresponds to the point where the congested zone (zone II)
and free-flow zone (zone III) meet, in agreement with traffic-flow models. From this plot
we see also that the flowrate calculated from the hydrodynamic relation $J= \rho V W$
(where W is the width of the bottleneck) is close to the flowrate calculated from the number
of refilling particles per second.  Thus our microscopic model reproduces the dependency
of the flowrate on the obstacle.

When the obstacle diameter is close to its optimal value, the behaviour of the flow at the 
bottleneck resembles the phase transition observed in vehicular traffic. To investigate this 
phase transition we plot mean velocity versus density for different snapshots in the simulations.  
Each dot in Fig.~\ref{fig:flowrate2} (d) represents  density  and  mean velocity at the bottleneck i
n one timeframe. We can recognize a phase transition between  congested and free-flow 
regimes at around $5.8$ particles per square meter. The best
fit of our velocity-density relation is a bilinear relation. This is in contrast with previous 
models without obstacles for pedestrian evacuation and gravity-driven outflow, which are generally based on  linear
~\cite{helbing97aac} or monotonically decreasing~\cite{treiber1999dpa} velocity-density relationships. We also note that this velocity-density relation with  two regimes does not capture the additional phase 
 transition from the congested to the hysteresis regime mentioned above. In the hysteresis regime the  
 velocity depends on particle diameter rather than density, which stays constant  around $7.3$ 
 particles per square meter.  A more sophisticated traffic-flow model is required to reproduce the full 
 phenomenology of flow of particles around an obstacle. Our phase transition differs also from that 
 observed in traffic flow in highways \cite{kerner97} where the flow, and not the mean velocity, peaks 
 at a critical density.

To conclude, the flowrate in a hopper with an obstacle placed before the bottleneck depends primarily 
on two parameters: the aperture, which is the minimum distance between the obstacle and the hopper, 
and the angle of the hopper. Depending on the aperture we detected three regimes: hysteresis flow, 
congested flow, and free flow. If the hopper angle is lower than $40^o$, the flowrate at the transition 
from congested to free flow is larger than the flowrate without  obstacle.  While this transition is similar to 
the one observed in traffic flow, the new transition from congested to hysteretic flow is not captured using a
 simple velocity-density relation. A microscopic model providing a complete analysis of this transition could 
 reap future benefits in increasing throughput and efficiency in a range of situations, from crowd control to 
 bulk-material transport in plants. 
Future investigations will be conducted to characterize the velocity field, in particular to study the transition from funnel flow to mass flow, and analyze the  temporal periodic fluctuations in the velocity field, which  are similar to the Karman oscillations observed when fluids flow around an obstacle.  


We thank M. Burd, N. Shiwakoti, A. Ramirez, I. Zuriguel, B. Pailthorpe, and J~.D Mu\~noz-Casta\~no for support and helpful discussions. Computations were performed on the Australian Earth Systems Simulator.
LMO is grateful for support from a University of Queensland ResTeach fellowship. FAM acknowledges the support of the Australian Research Council (project no. DP0772499).

\bibliographystyle{apsrev}
\bibliography{panicnew}

\end{document}